\documentclass[times,authoryear,fleqn,final,11pt]{article}		
\usepackage{bigints}
\usepackage{booktabs,stmaryrd}

\usepackage{graphicx}
\usepackage{psfrag,eepic,pstricks,rotating} 

\usepackage[font=small,labelfont=bf,nooneline]{caption}

\usepackage{amsmath,mathrsfs} 

\usepackage{amssymb}

\usepackage{lineno}
\usepackage[round]{natbib}
\bibpunct{[}{]}{;}{a}{,}{,}
\begin{document}

\begin{center}
{\huge \bf Thermal convection in a spherical shell\\ 
with melting/freezing at either or both of its boundaries}\\

\vspace*{0.5cm}
{\large Renaud Deguen$^*$}\\
\vspace*{0.5cm}
{Institut de M\'ecanique des Fluides de Toulouse,\\ Universit\'e de Toulouse (INPT, UPS) and CNRS.\\ All\' ee C. Soula, Toulouse, 31400, France}\\
{$^*$renaud.deguen@imft.fr}
\end{center}





\begin{abstract}

In a number of geophysical or planetological settings, including Earth's inner core, a silicate mantle crystallizing from a magma ocean, or an ice shell surrounding a deep water ocean -- a situation possibly encountered in a number of Jupiter and Saturn's icy satellites --, a convecting crystalline layer is in contact with a layer of its melt.
Allowing for melting/freezing at one or both of the boundaries of the solid layer is likely to affect the pattern of convection in the layer.
We study here the onset of thermal convection in a viscous spherical shell with dynamically induced melting/freezing at either or both of its boundaries. 
It is shown that the behavior of each interface -- permeable or impermeable -- depends on the value of a dimensional number $\mathcal{P}$ (one for each boundary), which is the ratio of a melting/freezing timescale over a viscous relaxation timescale.
A small value of $\mathcal{P}$ corresponds to permeable boundary conditions, while a large value of $\mathcal{P}$ corresponds to impermeable boundary conditions.
The linear stability analysis predicts a significant effect of semi-permeable boundaries when the number $\mathcal{P}$ characterizing either of the boundary is small enough: 
allowing for melting/freezing at either of the boundary results in the emergence of larger scale convective modes. 
The effect is particularly drastic when the outer boundary is permeable, since the degree 1 mode remains the most unstable even in the case of thin spherical shells. 
In the case of a spherical shell with permeable inner and outer boundaries, the most unstable mode consists in a global translation of the solid shell, with no deformation. 
In the limit of a full sphere with permeable outer boundary, this corresponds to the "convective translation" mode recently proposed for Earth's inner core.
As an example of possible application, we discuss the case of thermal convection in Enceladus' ice shell assuming the presence of a global subsurface ocean, and found that melting/freezing could have an important effect on the pattern of convection in the ice shell.

\end{abstract}

\section{Introduction}


The seismologically observed  hemispherical asymmetry  of the inner core \citep{Tanaka97,Niu01,irving2011} has recently been interpreted as resulting from a high-viscosity mode of  thermal convection, consisting in a translation of the inner core with melting on one hemisphere and solidification on the other \citep{Monnereau2010,Alboussiere2010}.
This "convective translation" regime can exist because the boundary between the inner core and the outer core is a phase change interface, which means that deforming the inner core boundary (ICB) by internal stresses can induce melting or freezing.
Melting occurs when the ICB is displaced outward, and crystallization occurs when the ICB is displaced inward, at a rate which depends on the ability of outer core convection to supply or evacuate the latent heat of phase change. 
Because there is no deformation, and therefore no viscous dissipation, associated with it,  the translation mode is dominant whenever phase change at the inner core boundary proceeds at a fast enough rate.



The situation where a convective crystalline shell is in contact with its melt 
 is encountered in 
  a number of other geophysical or planetological problems, including convection in a silicate mantle crystallizing from below from a magma ocean, or from a  basal magma ocean \citep{Labrosse2007,ulvrova2012}, or convection in an ice shell surrounding a deep water ocean, a situation possibly encountered in several of Jupiter and Saturn's  icy satellites  \citep{kivelson2000galileo,spohn2003,tyler2008strong}.
If one of the boundary is impermeable, the translation mode predicted for a full sphere obviously cannot exist, but we might anticipate that allowing for phase change at the other boundary will modify the pattern of convection and favor larger scale modes \citep{monnereau2002}. 

We will study here the onset of thermal convection in a uniformly heated spherical shell with boundary conditions allowing for dynamically induced melting or freezing at either or both of the boundary \citep{Deguen2013}.
The problem set-up and the system of equations are described in section \ref{SectionProblemDefinition}, with an emphasis on the formulation of the boundary conditions.
The steady basic solution  of the system of equations is found in section \ref{SectionBasicSolution}.
In section \ref{SectionStabilityAnalysis}, we perform a linear stability analysis of the set of equations described in section \ref{SectionProblemDefinition}, which allows to determine the pattern of the first unstable mode  as a function of the shell outer-to-inner radius ratio and of two non-dimensional numbers describing the resistance to phase change at each boundary.
The results and possible applications are discussed in section \ref{SectionResults}.

%


\section{Problem definition}
\label{SectionProblemDefinition}

\begin{figure}[t]
\begin{center}
\includegraphics[width=0.6\linewidth]{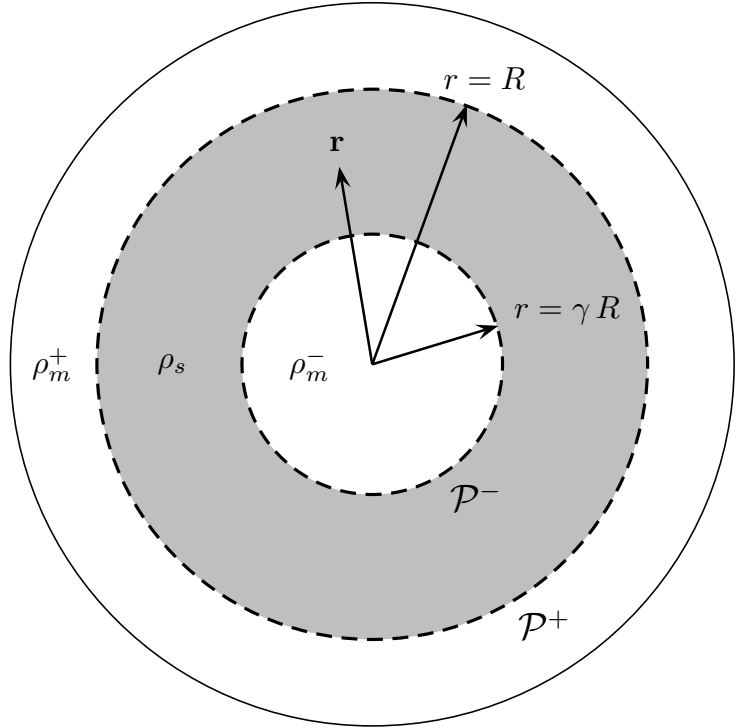}
\end{center}
\caption{A sketch of the problem considered here. 
\label{Sketch}
}
\end{figure}

We consider a viscous solid spherical shell of outer radius $R$ and inner radius $\gamma R$, in contact with melt layers either above or below, or both (see Figure \ref{Sketch}).
Superscripts "+" or "-" will be used for quantities taken at the outer or inner boundary, respectively. 
The solid shell has  constant density $\rho_s$, 
the layers below and above have densities $\rho_m^-$ and $\rho_m^+$, respectively, and we note $\Delta \rho^+=\rho_m^+-\rho_s$ and $\Delta \rho^-=\rho_m^--\rho_s$.
To insure long term mechanical stability of the solid layer, we must have $\rho_m^->\rho_s>\rho_m^+$, or $\Delta \rho^+>0$ and $\Delta \rho^-<0$.
The inner and outer boundaries are phase change interfaces, and melting and freezing can therefore occur when the interface is displaced by internal stresses. 
This will be described (section \ref{SectionBoundaryConditions}) with a parametrization of the relationship between the freezing or melting rate and the dynamic topography of the interface, which has been developed for the describing convection in Earth's inner core \citep{Alboussiere2010,Deguen2013}.

To be consistent with the assumption of constant density $\rho_s$, the acceleration of gravity $\mathbf{g}$ in the spherical shell is assumed to vary linearly with radius $r$, 
$\mathbf{g} = -g'\, r\, \mathbf{e}_r$ ,
where $g'=dg/dr=g^+/R=C^{st}$, 
which is relevant to situations where the depth dependence of the density is too small to have a significant effect on the mean gravity profile.
While this is not true in a number of situations of geophysical interest (like in Earth's mantle), we will make this assumption for two reasons: (i) it is (mathematically) the simplest configuration \citep{Chandrasekhar1961}, and (ii) the case of the inner core, for which $g$ is essentially linear in $r$,  corresponds to the limit $\gamma\rightarrow 0$ of the problem discussed here. 
Considering a more general form for $\mathbf{g}$ is likely to give qualitatively similar results.

The spherical shell is heated volumetrically at a rate $\rho_s c_{ps}S$ (with $S$ in K/s).
The rheology is assumed to be Newtonian and temperature and pressure independent, with a constant viscosity $\eta$. 
Thermal convection in the spherical shell is then described by the conservation equations for mass, momentum, and entropy, which take the form
\begin{linenomath}
\begin{align}
{\bf{\nabla} } \cdot {\bf u} &= 0, \label{continuity} \\
{\bf 0} &= - {\bf{\nabla} } p  -  {\alpha\, \rho_s}\, \Theta \, {\bf g} + {\eta } {\bf{\nabla} }^2 {\bf u} , \label{momentum_ALA2} \\
\frac{\partial \Theta}{\partial t} + \mathbf{u}\cdot \nabla \Theta &= \kappa {\nabla}^2\ \Theta + S, \label{entropy} 
\end{align}
\end{linenomath}
under the Boussinesq approximation.

\subsection{Boundary conditions}
\label{SectionBoundaryConditions}


The rate of melting/freezing at each interface depends on the ability of convective motion in the melt layer to transport the heat absorbed or released by the phase change. 
Given a topography $h(\theta,\phi)$ of the boundary\footnote{defined here in reference to the isopotential surface which coincides on average with the boundary.}, the rate of erosion of the topography by melting or freezing is set by a balance between the rate of latent heat release or absorption, $\rho_s L\, \partial h/\partial t$, with the convective heat flux on the melt side, which should scale as $\rho_l c_{pl}  u' \delta \Theta$, where $L$ is the latent heat of melting, $c_{pl}$ the specific heat capacity of the melt, $u'$ a typical velocity scale for convective motion in the melt layer, and $\delta \Theta(\theta,\phi)$  the difference of potential temperature between the boundary and the adjacent melt. 
The boundary is assumed to remain very close to thermodynamic equilibrium (more justifications in \cite{Deguen2013}), and is therefore at the melting temperature $T_m$.
The potential temperature variation $\delta \Theta$ along the boundary results from the combined effect of the pressure dependency of $T_m$ and of the adiabat in the melt layer, so that a topography $h$ induces a difference of potential temperature between the boundary and the melt layer given by 
\begin{linenomath}
\begin{equation}
\delta \Theta = - (m_p - m_\mathrm{ad}) \rho_l^\pm g^\pm  h, 
\end{equation}
\end{linenomath}
where $m_p = \partial T_s/\partial P$ is the Clapeyron slope, and $m_\mathrm{ad}={\partial T_\mathrm{ad}}/{\partial P}$ the adiabatic gradient in the melt layer.
With this expression for $\delta \Theta$, the heat balance  described above gives
\begin{linenomath}
\begin{equation}
u_r - \frac{\partial h}{\partial t} = \frac{h}{\tau_\phi^{\pm}},  \label{MeltingKinetics}
\end{equation}
\end{linenomath}
where the timescale for phase change, $\tau_\phi^{\pm}$, is 
\begin{linenomath}
\begin{equation}
\tau_\phi^{\pm} \sim \frac{\rho_s\, L}{\rho _l^2 c_{pl} \left| m_p - m_\mathrm{ad} \right| g^\pm u' }. \label{tauphi}
\end{equation}
\end{linenomath}
With $m_p=T_m \Delta\rho^\pm/(\rho_s \rho_l^\pm L)$ from the Clapeyron relation, Equation \eqref{tauphi} can be rewritten as
\begin{linenomath}
\begin{equation}
\tau_\phi^{\pm} \sim \frac{\rho_s^2\,   L^2}{\rho _l^\pm |\Delta\rho^\pm| c_{pl} T_m \left(1  - m_\mathrm{ad}/m_p \right) g^\pm u' }. \label{tauphi2}
\end{equation}
\end{linenomath}
Assuming that the phase-change timescale $\tau_\phi^{\pm}$ and the viscous relaxation timescale $\tau_\eta=\eta/(|\Delta \rho^\pm| g^\pm R)$ are both small compared to the dynamical timescale of the shell (overturn time), we can neglect $\partial h/\partial t$ in Eq. \eqref{MeltingKinetics}, which gives the boundary condition
\begin{linenomath}
\begin{equation}
u_r  = \frac{h}{\tau_\phi^{\pm}}.  \label{MeltingKinetics2}
\end{equation}
\end{linenomath}

The mechanical boundary conditions are tangential stress-free conditions and continuity of the normal stress at both boundary. 
Under the assumption of small topography, the stress-free tangential condition writes
\begin{linenomath}
\begin{align}
\tau _{r \theta} = \eta \left[ r \frac{\partial }{\partial r} \left( \frac{u_\theta }{r} \right) + \frac{1}{r } \frac{\partial u_r}{\partial \theta } \right] &= 0 , \label{stresstheta} \\
\tau _{r \phi} = \eta \left[ r \frac{\partial }{\partial r} \left( \frac{u_\phi }{r} \right) + \frac{1}{r \sin \theta } \frac{\partial u_r}{\partial \phi} \right] &= 0 , \label{stressphi} 
\end{align}
\end{linenomath}
at $r=\gamma$ and 1, where $\tau _{r \theta}$ and $\tau _{r \phi}$ are the $(r,\theta)$ and $(r,\phi)$ components of the deviatoric stress tensor  $\underline{\mathbf{\tau}}$.
Continuity of the normal stress at each boundary is written as
\begin{linenomath}
\begin{equation}
\llbracket \tau_{rr} - p \rrbracket = \left\llbracket 2 \eta \frac{\partial u_r}{\partial r} - p  \right\rrbracket _{h} = 0, \label{normalstress1}
\end{equation}
\end{linenomath}
where $\llbracket \dots \rrbracket$ denotes the difference of a quantity across the boundary.
When expanded around the mean position of the boundary, Eq. \eqref{normalstress1} gives
\begin{linenomath}
\begin{equation}
 - \Delta \rho^\pm  g\, {h}  - 2 \eta \frac{\partial u_r}{\partial r} + {p} = 0  \label{normalstress2}
\end{equation}
\end{linenomath}
under the assumption that pressure fluctuations on the melt side are negligible compared to pressure fluctuations on the solid side \citep[e.g.][]{Ribe07}. 
With $h$ related to $u_r$  by Eq. \eqref{MeltingKinetics2}, Eq. \eqref{normalstress2} gives a boundary condition for $u_r$ only:
\begin{linenomath}
\begin{equation}
 - \Delta \rho^\pm  g\,\tau_\phi^\pm\, u_r  - 2 \eta \frac{\partial u_r}{\partial r} + {p} = 0.
\end{equation}
\end{linenomath}
The topography $h$ is an implicit variable of the problem, and can be calculated \textit{a posteriori} from the radial velocity at the boundary.

\subsection{Non-dimensional set of equations}
\label{SectionEquationsSet}

The governing equations and boundary conditions are now made dimensionless using the thermal diffusion timescale $\kappa/R^2$, the outer radius $R$, $\kappa / R$, $\eta \kappa / R^2$ and $S R^2  / (6 \kappa )$ as scales for time, length, velocity, pressure and potential temperature, respectively. Using the same symbols for dimensionless quantities, the system of equations (\ref{continuity}-\ref{entropy}) is then written as
\begin{linenomath}
\begin{align}
{\bf{\nabla} } \cdot {\bf u} &= 0, \label{continuityND} \\
{\bf 0} &= - {\bf{\nabla} } p  +  Ra\, \Theta \, {\bf r} +  {\bf{\nabla} }^2 {\bf u} , \label{momentum_ALA2_ND} \\
  \frac{\partial \Theta}{\partial t} + \mathbf{u}\cdot \nabla \Theta &=  {\nabla}^2\ \Theta + 6, \label{entropy_ND} 
\end{align}
\end{linenomath}
where the Rayleigh number is defined as
\begin{linenomath}
\begin{equation}
Ra = \frac{\alpha \rho_s g^+ S R^{5}}{6 \eta \kappa^2}.
\end{equation}
\end{linenomath}
The Rayleigh number defined here is based on the outer radius $R$, not the shell thickness $(1-\gamma)R$.
Also, note that the Rayleigh number used here is half that defined by \cite{Chandrasekhar1961}.
The dimensionless boundary conditions at $r = \gamma$ or $ 1$ can be written
\begin{linenomath}
\begin{align} 
\Theta(\gamma)=\Theta_\gamma,\quad \Theta(1)&=0 , \label{bcTfa} \\
 r \frac{\partial }{\partial r} \left( \frac{u_\theta }{r} \right) + \frac{1}{r } \frac{\partial u_r}{\partial \theta }  = r \frac{\partial }{\partial r} \left( \frac{u_\phi }{r} \right) + \frac{1}{r \sin \theta } \frac{\partial u_r}{\partial \phi} &= 0 , \label{stressthetafa} \\
  \pm {\cal{P}^{\pm}}  u_r   + 2  \frac{\partial u_r}{\partial r} - {p} &= 0, \label{normalstressfa}   
\end{align}
\end{linenomath}
where the "phase change numbers" $\mathcal{P}^+$ and  $\mathcal{P}^-$ are defined as \citep{Deguen2012,Deguen2013}
\begin{linenomath}
\begin{equation}
\mathcal{P}^{\pm} = \frac{\tau _\phi^\pm}{\tau_\eta},   \label{P}
\end{equation}
\end{linenomath}
where $\tau _\phi^\pm$ is the timescale for erosion of a topography by melting or freezing, as defined in Eq. \eqref{tauphi}, and $\tau_\eta=\eta/(|\Delta \rho^\pm| g^\pm R)$ is the viscous relaxation timescale at the lengthscale $R$.
The phase change numbers $\mathcal{P}^+$ and $\mathcal{P}^-$ are measures of the resistance to phase change on each boundary.
In the limit of infinite $\mathcal{P}^\pm$, the boundary condition \eqref{normalstressfa} reduces to the condition $u_r=0$, which corresponds to impermeable conditions. 
In contrast, when $\mathcal{P}^\pm\rightarrow 0$, Eq. \eqref{normalstressfa} implies that the normal stress tends toward 0 at the boundary, which corresponds to fully permeable boundary conditions \citep{monnereau2002}. 
The general case of finite $\mathcal{P}^\pm$ gives boundary conditions for which the rate of phase change at the boundary (equal to $u_r$) is proportional to the normal stress induced by convection within the spherical shell.
Note that we have defined $\mathcal{P}^-$ and $\mathcal{P}^+$ using the absolute value of $\Delta \rho^\pm$, so that both $\mathcal{P}^-$ and $\mathcal{P}^+$ are positive. Because $\Delta \rho^-$ is negative, this introduces a minus sign before $\mathcal{P}^-$ in the boundary condition \eqref{normalstressfa} for the inner boundary.

With the assumptions made so far, the velocity field is known to be purely poloidal \citep{Ribe07}, and we introduce the poloidal scalar $P$ defined such that ${\bf u} = {\bf{\nabla} } \times {\bf{\nabla} } \times \left( P\, {\bf r} \right)$.
Taking the curl of the momentum equation \eqref{momentum_ALA2_ND} gives
\begin{linenomath}
\begin{equation}
Ra\,  L^2 \Theta = \left( {\nabla }^2 \right) ^2 L^2 P , \label{poloidalP}
\end{equation}
\end{linenomath}
where the angular momentum operator $L^2$ is
\begin{linenomath}
\begin{equation}
L^2 = -  \frac{1}{\sin \theta} \frac{\partial}{\partial \theta} \left( \sin \theta  \frac{\partial}{\partial \theta} \right) - \frac{1}{\sin ^{2} \theta}\frac{ \partial^2}{\partial \phi^2}.
\end{equation}
\end{linenomath}
Horizontal integration of the momentum equation \eqref{momentum_ALA2_ND} \citep{Ribe07} shows that, on both boundary,
\begin{linenomath}
\begin{equation}
- p + \frac{\partial }{\partial r } \left( r {\nabla}^2 P  \right) = C^{st}. \label{horizontal}
\end{equation}
\end{linenomath}
Using this  expression to eliminate $p$ in the boundary condition \eqref{normalstressfa}, and noting that $u_r = \frac{1}{r}  L^2 P$, 
continuity of the normal stress at each boundary  (equation \eqref{normalstressfa}) gives the following boundary condition for the poloidal scalar at $r=1$ or $\gamma$:
\begin{linenomath}
\begin{equation}
\frac{\partial }{\partial r} \left( r {\nabla}^2 P  - \frac{2}{r} L^2 P \right) - \pm  {\cal{P}^\pm}  \frac{L^2 P}{r}    = C^{st}, \label{normalstressf2a}
\end{equation}
\end{linenomath}
while the stress-free conditions (\ref{stressthetafa}) give
\begin{linenomath}
\begin{equation}
\frac{\partial^2 P}{\partial r^2} + \left( L^2 -{2} \right) \frac{P}{r^2} = C^{st},\ r=\gamma\ \mbox{or}\, 1  . \label{stressfreeP}
\end{equation}
\end{linenomath}

\section{Steady basic solution}
\label{SectionBasicSolution}

The governing equations and boundary conditions presented in section \ref{SectionEquationsSet} admit a steady solution (denoted by an overbar $\bar{...}$) in which the velocity field is $\mathbf{\bar u}=\mathbf{0}$ and the potential temperature field $\bar \Theta$ is given by the steady state, conductive version of Eq. \eqref{entropy}, which writes
\begin{linenomath}
\begin{equation}
0 = \nabla^2 \bar\Theta + 6.  \label{EqHeatConduction} 
\end{equation}
\end{linenomath}
With $\bar \Theta(r=1)=0$, the general solution of Eq. \eqref{EqHeatConduction} is of the form
\begin{linenomath}
\begin{equation}
\bar \Theta = a + \frac{1-a}{r} - r^2, \label{Tbase}
\end{equation}
\end{linenomath}
where the constant $a$ depends on the thermal boundary condition (imposed temperature or flux) at $r=\gamma$. 
The stability analysis could be carried out for the general potential temperature profile given by Eq. \eqref{Tbase}, but we will here consider only the case $a=1$.
This is mathematically simpler, and,  in addition, will allow us to extrapolate easily the results to the case of Earth's inner core, for which the basic diffusive potential temperature profile is given by $\bar \Theta = 1 - r^2$ \citep{Deguen2013}.
The potential temperature at $r=\gamma$ is $\Theta(\gamma)=1-\gamma^2$.

\section{Linear stability analysis}
\label{SectionStabilityAnalysis}

We now investigate the stability of the basic conductive state against infinitesimal perturbations of the temperature and velocity fields.
The present analysis follows the analysis presented in \cite{Chandrasekhar1961} (Chapter VI-60), where the stability analysis is treated in the case of impermeable boundaries, which corresponds to the limit of infinite $\mathcal{P}^-$ and $\mathcal{P}^+$. 
The case of thermal convection in a full sphere with boundary conditions as described above, which corresponds to the limit $\gamma \rightarrow 0$ of the problem considered here, has been treated in \cite{Deguen2013}. 



The temperature field is written as the sum of the conductive temperature profile given by Eq. \eqref{Tbase} and infinitesimal disturbances $\tilde \Theta$, $\Theta(r,\theta,\phi,t) = \bar \Theta(r) + \tilde \Theta(r,\theta,\phi,t)$. 
The velocity field perturbation is denoted by $\tilde{\mathbf{u}}(r,\theta,\phi,t)$, and has an associated poloidal scalar $\tilde P(r,\theta,\phi,t)$.
We expand the temperature and poloidal disturbances in spherical harmonics,
\begin{linenomath}
\begin{align}
\tilde \Theta &= \sum_{l=0}^\infty \sum_{m=-l}^l \tilde t_l^m(r) Y_l^m(\theta,\phi)\, \mathrm{e}^{\sigma_l t}, \label{TDecomposition} \\
\tilde P &=\sum_{l=1}^\infty \sum_{m=-l}^l \tilde p_l^m(r) Y_l^m(\theta,\phi)\, \mathrm{e}^{\sigma_l t}, \label{PDecomposition}
\end{align}
\end{linenomath}
where $\sigma_l$ is the growth rate of the degree $l$ perturbations (note that since $m$ does not appear in the  system of equations, the growth rate is function of $l$ only, not $m$). 

The only non-linear term in the system of equations is the advection of heat $\mathbf{u}\cdot \nabla \Theta$ in Equation \eqref{entropy_ND}, which is linearized as 
\begin{linenomath}
\begin{equation}
\tilde u_r \frac{\partial \bar \Theta}{\partial r} = -2 r \tilde u_r = - 2 L^2 \tilde P .  
\end{equation}
\end{linenomath}
The resulting linearized transport equation for the potential temperature disturbance is
\begin{linenomath}
\begin{equation} 
\left(\frac{\partial }{\partial  t} -  {\nabla}^2 \right)\tilde \Theta=  2 L^2 \tilde P + 6.
\end{equation} 
\end{linenomath}
Using the decompositions \eqref{TDecomposition} and \eqref{PDecomposition}, the linearized system of equations is then, for $l\geq 1$,
\begin{linenomath}
\begin{align}
Ra\, \tilde t_l^m &= \mathcal{D}_l^2 \tilde p_l^m, \label{MomentumMarg} \\
\left( \sigma_l -  \mathcal{D}_l \right) \tilde t_l^m &= 2 l (l+1) \tilde p_l^m ,  \label{HeatMarg}
\end{align}
\end{linenomath}
with the stress-free boundary condition written as
\begin{linenomath}
\begin{equation}
\frac{d^2 p_l^m }{d r^2} + \left[ l(l+1) -2 \right] \frac{p_l^m}{r^2} = 0, \ \ \ \ l \geq 1, \label{stressfreePlm}   
\end{equation}
\end{linenomath}
with $r=1$ or $\gamma$ on the upper or lower boundaries, and the boundary conditions derived from the continuity of the normal stress given by
\begin{linenomath}
\begin{align}
\frac{d }{d r} \left(r \mathcal{D}_l p_l^m -2l(l+1) \frac{p_l^m}{r}   \right) &= + l(l+1) {\cal{P}^{+}} (t) \frac{p_l^m}{r} , \label{normalstressPlm_plus} \\
\frac{d }{d r} \left(r \mathcal{D}_l p_l^m -2l(l+1) \frac{p_l^m}{r}   \right) &= - l(l+1) {\cal{P}^{-}} (t) \frac{p_l^m}{r}, \label{normalstressPlm_moins}
\end{align}
\end{linenomath}
at the outer and inner boundaries, respectively (note the different signs of the right-hand-side terms).
The operator $\mathcal{D}_l$ is defined as
\begin{linenomath}
\begin{equation}
\mathcal{D}_l = \frac{d^2 }{d r^2} + \frac{2}{r} \frac{d}{d r} - \frac{l(l+1)}{r^2}. \label{Dl}
\end{equation}
\end{linenomath}


\begin{figure}[t]
\begin{center}
\includegraphics[width=0.5\linewidth]{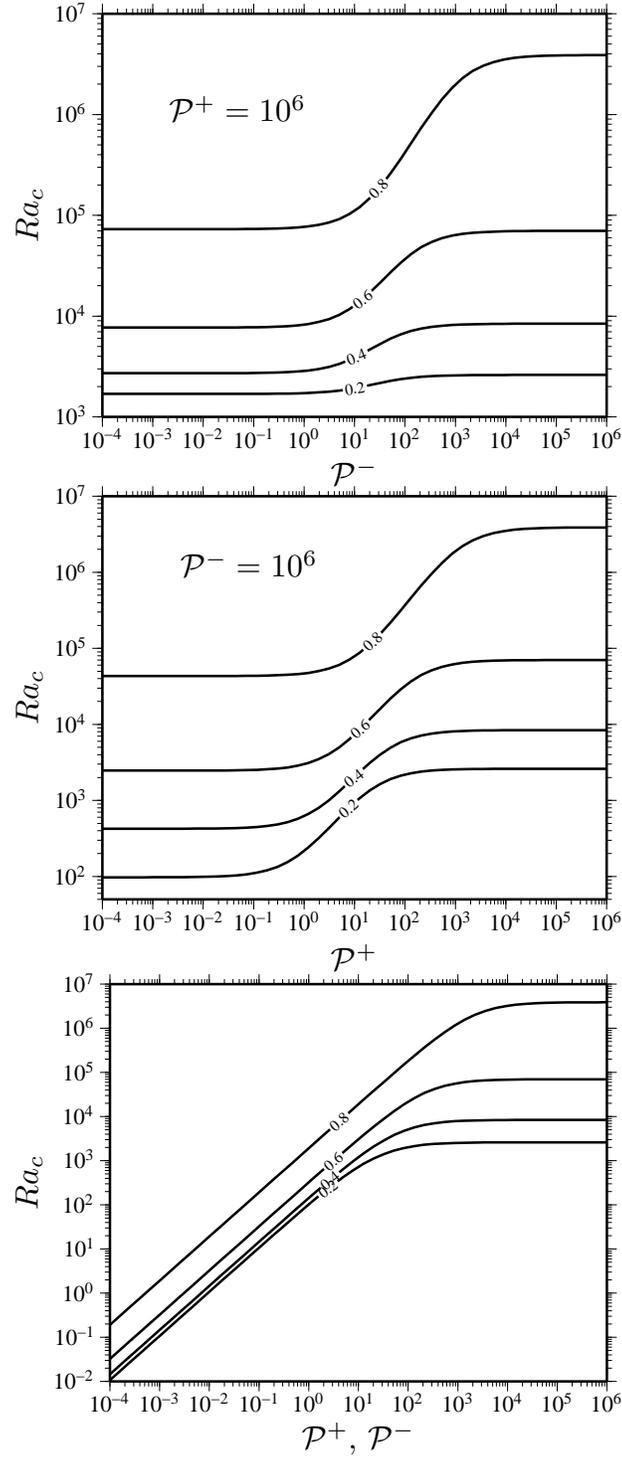}
\end{center}
\caption{Critical Rayleigh number of the $l=1$ mode for : \textbf{a)} impermeable outer boundary, and variable $\mathcal{P}^-$, 
\textbf{b)} impermeable inner boundary, and variable $\mathcal{P}^+$,
\textbf{c)} variable $\mathcal{P}^-$ and $\mathcal{P}^+$, with $\mathcal{P}^+=\mathcal{P}^-$.
\label{FigVariableP}
}
\end{figure}

We expand the potential temperature perturbation $\tilde t_l^m(r)$ as 
\begin{linenomath}
\begin{equation}
\tilde t_l^m = \frac{1}{\sqrt{r}} \sum_j A_{lj}\, \mathcal{C}_{ll}(\alpha_{lj} r), \label{Texp}
\end{equation}
\end{linenomath}
where the functions $ \mathcal{C}_{ll}(\alpha_{lj} r)$ are defined as 
\begin{linenomath}
\begin{equation}
\mathcal{C}_{ll}(\alpha_{lj} r) = J_{-\left(l+\frac{1}{2}\right)}(\alpha_{lj} \gamma)  J_{l+\frac{1}{2}}(\alpha_{lj} r) - J_{l+\frac{1}{2}}(\alpha_{lj} \gamma)  J_{-\left(l+\frac{1}{2}\right)}(\alpha_{lj} r) 
\end{equation}
\end{linenomath}
 \citep{Chandrasekhar1961}. Here $J_k$ denotes the  Bessel function of the first kind of degree $k$, and the constants $\alpha_{lj}$ are the $j$th zeros of the function $\mathcal{C}_{ll}(r)$.
By construction, $\mathcal{C}_{ll}(\alpha_{lj} \gamma)=0$.
As discussed by \cite{Chandrasekhar1961}, the functions $\mathcal{C}_{ll}(\alpha_j r)$ form an integral set of functions satisfying the orthogonality relation
\begin{linenomath}
\begin{equation}
\int_\gamma^1 \mathcal{C}_{ll}(\alpha_{lj} r) \mathcal{C}_{ll}(\alpha_{lk} r) r dr = N_{l+\frac{1}{2},j} \delta_{jk},
\end{equation}
\end{linenomath}
where
\begin{linenomath}
\begin{equation}
N_{l+\frac{1}{2},j} = \frac{2}{\pi^2 \alpha_{lj}^2}  \left[ \frac{J_{l+\frac{1}{2}}^2(\alpha_{lj} \gamma)}{J_{l+\frac{1}{2}}^2(\alpha_{lj})} - 1  \right].
\end{equation}
\end{linenomath}

Writing the poloidal scalar perturbations $\tilde p_l^m$ as
\begin{linenomath}
\begin{equation}
\tilde p_l^m(r) = \sum_j A_{lj}\, p_{lj}(r)  \label{Pexp}
\end{equation}
\end{linenomath}
and injecting the expansions of $\tilde t_l^m(r)$ and $\tilde p_l^m(r)$ given by Eq. \eqref{Texp} and \eqref{Pexp} in the momentum equation \eqref{MomentumMarg}, 
the functions $p_{lj}$ are solutions of the equation
\begin{linenomath}
\begin{equation}
 \mathcal{D}_l^2 p_{lj} = Ra\,  \frac{\mathcal{C}_{ll}(\alpha_{lj} r)}{\sqrt{r}}  . \label{MomentumMarg2}
\end{equation}
\end{linenomath}
Noting that  
\begin{linenomath}
\begin{equation}
\mathcal{D}_l \frac{\mathcal{C}_{ll}(\alpha_{lj} r)}{\sqrt{r}} = - \alpha_{lj}^2\, \frac{\mathcal{C}_{ll}(\alpha_{lj} r)}{\sqrt{r}},
\end{equation}
\end{linenomath}
equation \eqref{MomentumMarg2} has a general solution of the form
\begin{linenomath}
\begin{equation}
p_{lj} = \frac{Ra}{\alpha_{lj}^4} \frac{ \mathcal{C}_{ll}(\alpha_{lj} r)}{\sqrt{r}} + B_1^{j} r^l + B_2^{j} r^{l+2} + B_3^{j} r^{-(l+1)} + B_4^{j} r^{-(l-1)}.   \label{GeneralSolution_p} 
\end{equation}
\end{linenomath}
The coefficients $B_{1...4}^{j}$ are determined by the boundary conditions at the inner and outer boundaries of the shell, as explained in Appendix \ref{AppendixBs}.

Injecting the above solution for $p_{lj}$ and the potential temperature expansion \eqref{Texp} in the linearized heat equation \eqref{HeatMarg}, we obtain after some manipulation an infinite set of linear equations in $A_{lj}$  (see \cite{Chandrasekhar1961}), which admits a non trivial solution only if its determinant is equal to zero. 
With our choice of basic state and $g \propto r$, and following \cite{Chandrasekhar1961}, we find a characteristic equation of the form
\begin{linenomath}
\begin{equation}
\left|\left| N_{l+\frac{1}{2},k}\left[ \frac{\alpha_k^2}{l(l+1) 2} - \frac{Ra}{\alpha_k^4}  \right] \delta_{kj} - Q_{kj}  \right|\right| = 0,  \label{CharacteristicEquation}
\end{equation}
\end{linenomath}
where $||...||$ denotes the determinant, and where the functions $Q_{kj}$ are defined as
\begin{linenomath}
\begin{equation}
Q_{kj}  = \int_\gamma^1\!\! \mathcal{C}_{ll}(\alpha_{k} r) \left[ B_1^{j} r^l + B_2^{j} r^{l+2} + B_3^{j} r^{-(l+1)} + B_4^{j} r^{-(l-1)} \right] r^{\frac{3}{2}} dr.
\end{equation}
\end{linenomath}
Solving Eq. \eqref{CharacteristicEquation} with the $B_{1...4}^{i}$  determined by the boundary conditions (Appendix \ref{AppendixBs}) gives the critical Rayleigh number $Ra_c^l$ for a perturbation of degree $l$.
The pattern of the first unstable modes can be calculated by solving the system  in ${A}_{lj}$ for given $\mathcal{P}^-$, $\mathcal{P}^+$ and $Ra$, which then allows to calculate the poloidal scalar $\tilde{p}_l^m$ from equations \eqref{Pexp} and \eqref{GeneralSolution_p}.


\section{Results and applications}
\label{SectionResults}

\begin{figure*}[ht]
\begin{center}
\includegraphics[width=\linewidth]{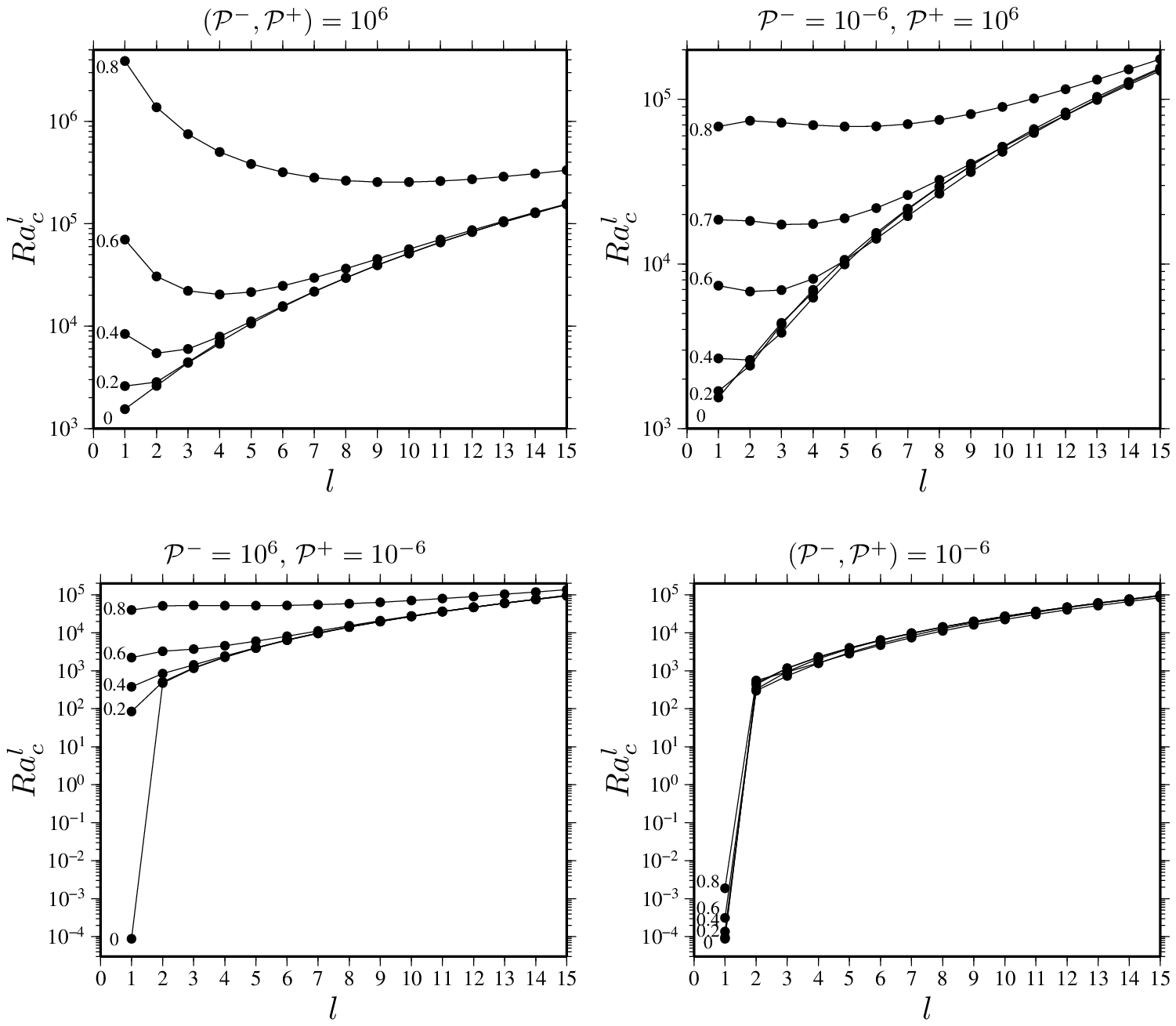}
\end{center}
\caption{Critical Rayleigh number for convection as a function of degree $l$, for the four end-member cases described in the text, for various values of $\gamma$. 
Note the different scales used for $Ra_c^l$.  
\label{FigureCriticalRa}
}
\end{figure*}

\begin{figure*}[ht]
\begin{center}
\includegraphics[width=\linewidth]{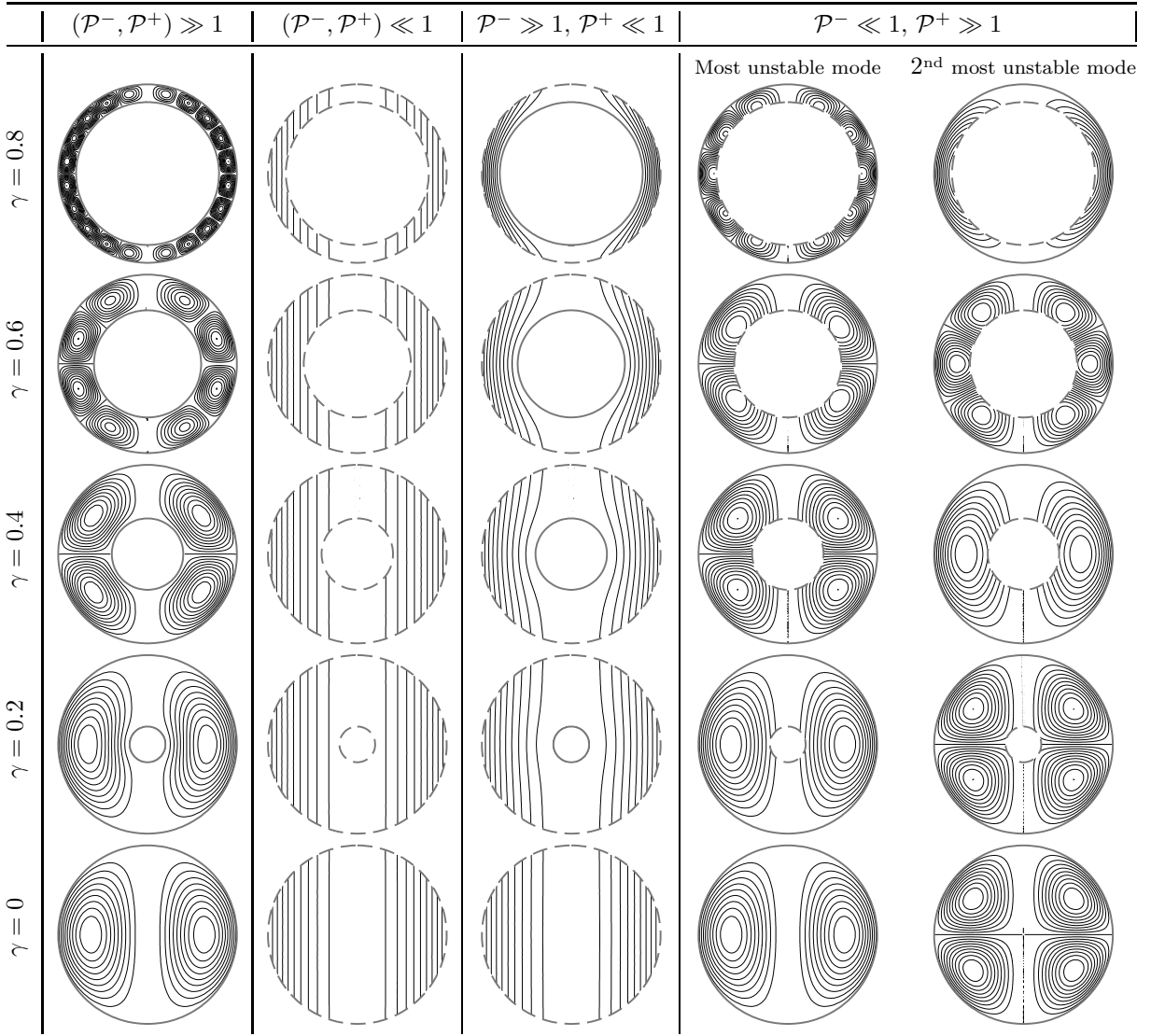}
\end{center}
  \caption{Pattern of the first unstable mode of thermal convection in a spherical shell (streamlines), with aspect ratio $\gamma=0$, 0.2, 0.4, 0.6 and 0.8, and $\mathcal{P}^-$ and $\mathcal{P}^+$ either small or large compared to 1. Impermeable boundaries ($\mathcal{P}^\pm\gg1$) are shown by a thick  line, permeable boundaries ($\mathcal{P}^\pm\ll 1$) are shown by a thick  dashed line. In the case $\mathcal{P}^-\ll 1$, $\mathcal{P}^+\gg 1$, we also show the second most unstable mode. Only the $m=0$ modes are shown.
\label{FigurePattern}
  }
\end{figure*}

\begin{figure}[t]
\begin{center}
\includegraphics[width=0.7\linewidth]{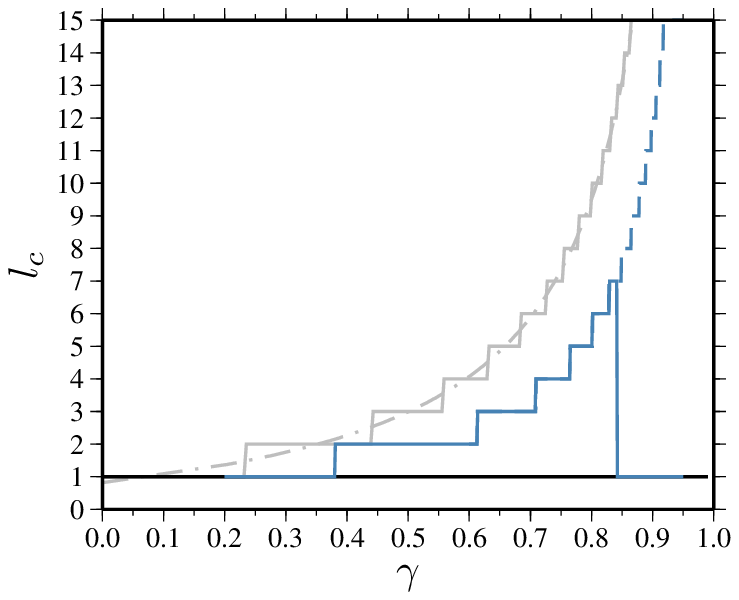}
\end{center}
\caption{Degree $l_c$ of the first unstable mode as a function of the aspect ratio $\gamma$, for different configurations.
The solid gray line corresponds to the case of impermeable inner and outer boundaries.
The solid blue line corresponds to the case of impermeable outer boundary and fully permeable inner boundary. 
The dashed blue line shows the degree of the local minimum at $l$ strictly larger than 1 in the case of impermeable outer boundary and fully permeable inner boundary (see text).
The solid black line corresponds to the cases of fully permeable inner and outer boundaries, and of fully permeable outer boundary and impermeable inner boundary, for which the most unstable mode is always the degree 1 mode. 
\label{DegreeFirstUnstableMode}
}
\end{figure}


Figure \ref{FigVariableP} shows the critical Rayleigh number corresponding to the degree one mode as a function of $\mathcal{P}^-$ and $\mathcal{P}^+$ for three configurations : (i) the outer boundary is impermeable ($\mathcal{P}^+\gg 1$) and $\mathcal{P}^-$ is varied from permeable to impermeable conditions; (ii) the inner boundary is impermeable ($\mathcal{P}^-\gg 1 $) and $\mathcal{P}^+$ is varied from permeable to impermeable conditions; and (iii) $\mathcal{P}^+=\mathcal{P}^-$, with boundary conditions varied from permeable to impermeable. 
For all three configurations, there is a marked change in the critical Rayleigh number  at some transitional value of $\mathcal{P}^-$ or $\mathcal{P}^+$, with the critical Rayleigh number being significantly smaller when $\mathcal{P}^-$ or $\mathcal{P}^+$ are smaller than this transitional value, corresponding to permeable conditions.

In what follow, we will focus  on end-members cases, for which each boundary is either permeable ($\mathcal{P}^\pm\ll1$) or impermeable ($\mathcal{P}^\pm\gg1$), which gives four end-member configuration. 
Figure \ref{FigureCriticalRa} shows the critical Rayleigh number as a function of $l$ and $\gamma$ for the four end-member cases.
The pattern of the first unstable mode (as well as the second for the $\mathcal{P}^-\ll 1$, $\mathcal{P}^+\gg 1$ cases) are shown in Figure \ref{FigurePattern}.
The degree of the first unstable mode is shown in Figure \ref{DegreeFirstUnstableMode} as a function of $\gamma$ for the four end-member cases.
Each end-member case is described below.

\subsection{$(\mathcal{P}^+,\mathcal{P}^-) \gg 1$ -- impermeable boundaries}


Letting $\mathcal{P}^+$ and $\mathcal{P}^-$ tend toward infinity, the problem tends toward the case of Rayleigh-B\'enard convection in a spherical shell with impermeable stress free boundaries, as discussed in \cite{Chandrasekhar1961}, and will be used as a reference case for the present study. 
The results found here are identical to that found by \cite{Chandrasekhar1961} (except that, as explained above,  the Rayleigh numbers shown here are half  that found by \cite{Chandrasekhar1961} because of different definitions).
The degree one mode is the first unstable mode for $\gamma$ smaller than $\simeq 0.23$. 
The degree of the first unstable mode then increases rapidly when $\gamma$ is increased (Figure \ref{DegreeFirstUnstableMode}). 
The corresponding wavelength is commensurate with the shell thickness $1-\gamma$:  
assuming a relationship of the form $l_c = a/(1-\gamma)+b$ (which, given that $\lambda_c \sim 1/l_c$ when $l_c\gg 1$, is equivalent to $\lambda_c \sim 1-\gamma$), least square inversion of $\ell_c(\gamma)$  gives $l_c=2.17/(1-\gamma) - 1.35$, which is shown as a grey dash-dotted line in Figure \ref{DegreeFirstUnstableMode}. 
The fit is indeed good, consistent with the assumption of a critical wavelength proportional to the layer thickness.

\subsection{$(\mathcal{P}^+,\mathcal{P}^-) \ll 1$ -- permeable inner and outer boundaries}

\begin{figure}
\begin{center}
\includegraphics[width=0.7\linewidth]{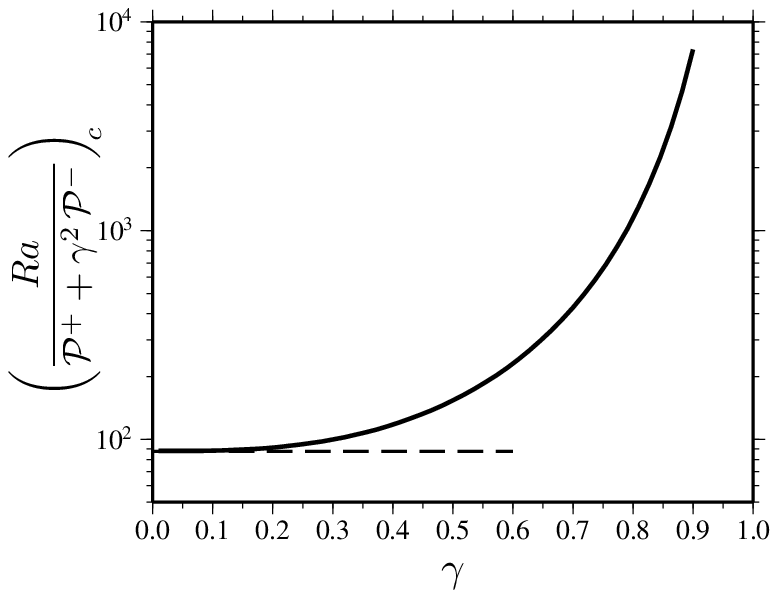}
\end{center}
\caption{Critical value of the quantity $\left(\frac{Ra}{{\mathcal{P}^+} + \gamma^2\, \mathcal{P}^-}\right)_c$ for the translation mode in the limit of small $\mathcal{P}^-$ and $\mathcal{P}^+$, as a function of the inner to outer radius ratio $\gamma$ (solid black line). 
The dashed black line shows the value of $175/2$ found analytically by \cite{Deguen2013} for a full sphere ($\gamma=0$).
\label{FigCriticalRaP}
}
\end{figure}

On the other extreme, when both boundaries are fully permeable, the first unstable mode is always the degree one mode (Figure \ref{FigureCriticalRa} and \ref{DegreeFirstUnstableMode}), which takes the form of a solid translation of the spherical shell  (Figure \ref{FigurePattern} and Appendix \ref{AppendixTranslation}).   
The limit of a full sphere ($\gamma=0$) corresponds to the "convective translation" mode recently put forward for Earth's inner core \citep{Monnereau2010,Alboussiere2010}.

Since this mode consists of a pure translation, there is no deformation, and therefore no viscous dissipation in the inner core.
This of course does not mean that this is a non-dissipative  mode. 
There is  viscous (and magnetic in the case of Earth's inner core) dissipation in the melt layer associated with the redistribution of the latent heat of phase change.
The melt layer must provide mechanical work to account for the dissipation associated with the redistribution of the latent heat, which means that this mode of convection is ultimately limited by the vigor of convective motions in the melt layer.

It can be shown (Appendix \ref{AppendixTranslation}) that the emergence of the translation mode requires that the quantity 
\begin{linenomath}
\begin{equation}
R_\mathcal{P} = \frac{Ra}{{\mathcal{P}^+} + \gamma^2\, \mathcal{P}^-}
\end{equation}
\end{linenomath}
is higher than a critical value which is a function of $\gamma$ only.
The quantity ${\mathcal{P}^+} + \gamma^2\, \mathcal{P}^-$ is, save for a factor $1+\gamma^2$, the boundary area weighted mean of $\mathcal{P}^-$ and $\mathcal{P}^+$. 
Figure \ref{FigCriticalRaP} shows the critical value of $R_\mathcal{P}$ for the translational instability as a function of $\gamma$ calculated using Eq. \eqref{CriticalRaP} of Appendix \ref{AppendixTranslation}. 
When $\gamma \rightarrow 0$, the critical value tends toward the  value of $(Ra/\mathcal{P}^+)_c=175/2=87.5$ found by \citet{Deguen2013} for a full sphere.
$R_\mathcal{P}$ then increases with $\gamma$.

The limit $\gamma\rightarrow 0$ is relevant for Earth's inner core dynamics \citep{Monnereau2010,Alboussiere2010,Mizzon2013,Deguen2013}. 
The case of a spherical shell with phase change at both boundaries 
might be relevant for the early dynamics of Earth's mantle, which may have start  crystallizing at mid-depth from a magma ocean, with a surface magma ocean and a basal magma ocean \citep{Labrosse2007}.

\subsection{$\mathcal{P}^+\ll 1$ and  $\mathcal{P}^- \gg 1$ -- permeable outer boundary, impermeable inner boundary}

When the inner boundary is impermeable ($\mathcal{P}^- \gg 1$) and the outer boundary fully permeable ($\mathcal{P}^+\ll 1$), the degree one mode is again found to be always the most unstable mode (Figs. \ref{FigureCriticalRa} and \ref{DegreeFirstUnstableMode}), even when $\gamma$ approaches 1.
In contrast with the case where both boundary are permeable, the degree 1 mode now does involve deformation,  and the decrease in $Ra_c$ compared to impermeable boundaries conditions is therefore not as drastic as when both boundary are permeable.
The critical Rayleigh number tends toward a finite value when $\mathcal{P}^+ \rightarrow 0$, because even with a fully permeable boundary, viscous dissipation always limit the development of the mode.

This configuration could be relevant for the initiation of convection in a silicate mantle crystallizing from below from a  magma ocean.
The stability analysis predicts that in this configuration the first unstable mode is the degree one mode shown in Figure \ref{FigurePattern}.
However, one key point in this configuration is the lifetime of the magma ocean. 
Melting/freezing at the interface would play a role only if the instability growth is fast enough compared to magma ocean crystallization, which can happen on a ky timescale  in the absence of an insulating atmosphere or crystallized lid \citep{solomatov2000}. 

\subsection{$\mathcal{P}^+\gg 1$ and  $\mathcal{P}^- \ll 1$ -- impermeable outer boundary, permeable inner boundary}

When the outer boundary is impermeable ($\mathcal{P}^+\gg 1$) and the inner boundary fully permeable ($\mathcal{P}^- \ll 1$), the relationship between the degree $l_c$ of the most unstable mode and $\gamma$ becomes non-monotonic (Figure \ref{DegreeFirstUnstableMode}, solid blue line).
$l_c$ first increases with $\gamma$, similarly to the case of impermeable boundaries (except $l_c$ is smaller when the inner boundary is permeable), but the $l=1$ mode becomes again the most unstable mode when $\gamma$ exceeds $\simeq 0.841$.
Looking at the critical Rayleigh number as a function of $l$ (Figure \ref{FigureCriticalRa}), there appears to be two local minima, one at $l=1$ and the other at a higher $l$, once $\gamma$ is larger than $\simeq 0.72$.
The two minima are quite close for all values of $\gamma$, which suggest that the $l=1$ mode would be important even if it is not the most unstable mode.
In Figure \ref{DegreeFirstUnstableMode}, we show in blue the degree of the most unstable mode (blue solid  line) in this configuration, as well as the degree of the local minimum at $l$ strictly larger than 1 (blue dashed line).

We show in Figure \ref{FigurePattern} the pattern of both the most  unstable and second most unstable modes.
The pattern of the degree one mode is found to be close to a truncated version of the pattern of the degree one mode of convection in a full sphere (compare with the $\gamma=0$ case). 

This configuration may be relevant for the dynamics of icy satellites having an ice mantle overlying a global subsurface water ocean, which might be the case of several of Jupiter and Saturn' moons, including Enceladus \citep{nimmo2006b,waite2009liquid}, Europa \citep{tyler2008strong}, Callisto, Ganymede and Titan \citep{spohn2003}.
It might also be relevant for thermal convection in a  silicate mantle overlying a basal magma ocean, as might have been the case on Earth early in its history \citep{Labrosse2007,ulvrova2012}.
The stability analysis suggest that the length scale of convection would be significantly larger if melting/freezing at the interface is important.

\section{Discussion and conclusions}

The linear stability analysis presented here predicts a significant effect of semi-permeable boundaries when either $\mathcal{P}^-$ or $\mathcal{P}^+$ are small enough: 
allowing for melting/freezing at either of the boundary results in the emergence of larger scale convective modes. 
The effect is particularly drastic when the outer boundary is permeable, since the degree 1 mode remains the most unstable even in the case of thin spherical shells. 
It seems likely that  allowing for melting/freezing at one boundary will still result in larger scale convection at supercritical conditions, but the results presented here  will clearly have to be supplemented by finite amplitude numerical calculations 
at supercritical conditions.
In addition, the assumption of Newtonian rheology and constant viscosity limits the direct applicability of our results.  
The effect of variable viscosity would have to be investigated, in particular for application to icy moons, for which order of magnitude variations of viscosity across the layer may be expected.
The pattern of convection is also likely to depend  on the temperature profile of the basic state \citep{mcnamara2005degree}. 

At this stage, we have suggested some possible geophysical or planetological applications of our results, but specific studies will be needed to assess the applicability of our results in particular geophysical or planetological settings. 
In each situation, the value of $\mathcal{P}$ of the boundary must be evaluated, which necessitates some understanding of the dynamics of the melt layer in contact with the solid layer. 

As an example, let us discuss  the case of Enceladus. 
Enceladus exhibits a strong hemispherical asymmetry, with the Southern hemisphere being much younger and active that the Northern hemisphere \citep{porco2006cassini}.
One plausible explanation for the observed asymmetry is degree one convection  \citep{grott2007degree,stegman2009}.
Enceladus may have a global subsurface ocean \citep{nimmo2006b,waite2009liquid,tyler2009ocean}, and it is therefore legitimate to consider the possible dynamical effect of melting/freezing at the inner boundary of the ice shell.
Whether phase change at the inner boundary of the ice shell can alter significantly the pattern of convection  depends on the value of $\mathcal{P}^-$: 
with $\gamma=0.6$ \citep{schubert2007enceladus}, the effect of phase change would be significant if  $\mathcal{P}^-$ is smaller than about $10$ (Figure \ref{FigVariableP}).
With a viscosity of order $10^{14}$ Pa~s (which corresponds to the viscosity near the melting point), a radius $R=250$ km, $|\Delta \rho^-|\simeq 50$ kg.m$^{-3}$ and $g^-\simeq 0.1$ m.s$^{-2}$, we find that $\mathcal{P}^-\lesssim 10$ if the timescale for phase change $\tau_\phi^-$ is smaller than about 25 years.
With $\tau_\phi^-$ given by Eq. \eqref{tauphi2}, $L=300$ kJ.kg$^{-1}$, $c_{pl}=4000$ K.kg$^{-1}$.K$^{-1}$, $T_s=275$ K, and $1-m_\mathrm{ad}/m_p \sim 1$,  this would require typical convective velocities around $\sim 1$ cm.s$^{-1}$ in the melt layer.
\cite{tyler2009ocean} estimates that eccentricity tides would have typical flow ampitude around $\sim 1$ mm.s$^{-1}$ in a $\sim 100$ km thick ocean and $\sim 1$ cm.s$^{-1}$ in a $\sim 10$ km thick ocean. 
This would give $\mathcal{P}^-$  in the range $10-10^2$, so $\mathcal{P}^-$ may plausibly be small enough for a significant effect of melting/freezing on the pattern of convection.
Including the effect of temperature on viscosity is likely to make the effect of melting/freezing stronger  because the effective viscosity for relaxation of a large scale topography would be larger, possibly by several order of magnitude, than the high homologous temperature value of $10^{14}$ Pa.s assumed here.  
This would yield a lower effective value of $\mathcal{P}^-$, and a more permeable boundary.
Whether or not the effect is strong enough to allow the emergence of a strong degree one convection mode remains an open question.
The answer might also depend in part of the dynamical effect of radial viscosity variations in the ice shell \citep{zhong2001degree,mcnamara2005degree}, which will have to be taken into account.


\section*{Acknowledgments}

This study was to a large extent motivated by Shijie Zhong's presentation on degree one structures in planetary mantles, given at the 2012 Core Dynamics workshop in Wuhan.
I would like to thank Dave Yuen and the organizing committee for the invitation, and Shijie Zhong for discussion.  
I gratefully acknowledge support from  grant ANR-12-PDOC-0015-01 of the ANR (Agence Nationale de la Recherche).

\appendix

\section{Coefficients $B_{1...4}^j$}
\label{AppendixBs}

The coefficients $B_{1...4}^{j}$ introduced in Eq. \eqref{GeneralSolution_p} are determined for each degree $l$ by the boundary conditions at the inner and outer boundaries of the shell.

Using expression \eqref{GeneralSolution_p} for $p_{lj}$, the tangential stress boundary condition (Eq. \eqref{stressfreeP})  gives
\begin{linenomath}
\begin{equation}
\begin{split}
 &B_1^j  \gamma^{l-2} (  l^2 - 1 ) 
+B_2^j \gamma^{l} l(l + 2) \\
+&B_3^j \gamma^{-l-3} l(l+2  ) 
+B_4^j \gamma^{-l-1} (   l^2 - 1 ) \\
=& Ra \frac{ \mathcal{C}'_{ll}(\alpha_{lj} \gamma)}{\alpha_{lj}^3{\gamma^{\frac{3}{2}}}},   \label{StressFreeGamma}
\end{split}
\end{equation}
\end{linenomath}
at  $r=\gamma$ and 
\begin{linenomath}
\begin{equation}
B_1^j   (  l^2 - 1 ) 
+B_2^j l(l + 2) 
+B_3^j  l(l+2  ) 
+B_4^j  (   l^2 - 1 ) 
= Ra \frac{ \mathcal{C}'_{ll}(\alpha_{lj})}{\alpha_{lj}^3},  \label{StressFree1}
\end{equation}
\end{linenomath}
at $r=1$.

The boundary condition \eqref{normalstressf2a} derived from the continuity of the normal stress gives 
\begin{linenomath}
\begin{equation}
\begin{split}
&B_1^{j} \gamma^{l-2} \left( 1 - l + \frac{ \cal{P}^{-}}{2}\gamma \right) 
+B_2^{j} \gamma^{l} \left( \frac{3}{l}  - l   + 1 +  \frac{ \cal{P}^{-}}{2}\gamma \right) \\
+ &B_3^{j} \gamma^{-l-3}\left( l + 2 +  \frac{ \cal{P}^{-}}{2}\gamma \right) 
+ B_4^{j} \gamma^{-l-1}\left( \frac{2l-1}{l+1} + l  +  \frac{ \cal{P}^{-}}{2}\gamma \right) \\
=  &\left[1 + \frac{\alpha_{lj}^2 \gamma^{2} }{2 l(l+1)}  \right]  Ra\, \frac{\mathcal{C}'_{ll}(\alpha_{lj} \gamma) }{\alpha_{lj}^3 \gamma^{\frac{3}{2}}}  \label{NormalStressGamma}
\end{split}
\end{equation}
\end{linenomath}
at $r=\gamma$ and
\begin{linenomath}
\begin{equation}
\begin{split}
&B_1^{j}  \left(  1 - l - \frac{ \cal{P}^{+}}{2} \right) 
+B_2^{j} \left( \frac{3}{l}  - l   + 1 -  \frac{ \cal{P}^{+}}{2} \right) \\
+ &B_3^{j} \left( l + 2 -  \frac{ \cal{P}^{+}}{2} \right) 
+ B_4^{j} \left( \frac{2l-1}{l+1} + l  -  \frac{ \cal{P}^{+}}{2} \right) \\
=  &\left[1 + \frac{\alpha_{lj}^2  }{2 l(l+1)}  \right]  Ra\, \frac{\mathcal{C}'_{ll}(\alpha_{lj} ) }{\alpha_{lj}^3 }   \label{NormalStress1}
\end{split}
\end{equation}
\end{linenomath}
at $r=1$.

Eqs. \eqref{StressFreeGamma}, \eqref{StressFree1}, \eqref{NormalStressGamma} and \eqref{NormalStress1} form a linear system of equations for $B_{1...4}^{j}$ which is solved for each degree $l$.
The $B_{1...4}^{j}$ are then used to calculate the functions $Q_{kj}$ in the characteristic equation \eqref{CharacteristicEquation}.

\section{Translation mode}
\label{AppendixTranslation}

We consider here the onset of the degree 1 mode in the limit of small $Ra$, $\mathcal{P}^-$, and $\mathcal{P}^+$, but finite $Ra/({\mathcal{P}^++\gamma^2 \mathcal{P}^-})$. 
With $l=1$, the tangential stress free conditions \eqref{StressFreeGamma} and \eqref{StressFree1} give 
\begin{linenomath}
\begin{align}
B_2^j   + B_3^j \gamma^{-5}  &= Ra \frac{ \mathcal{C}'_{11}(\alpha_{lj} \gamma)}{3\,\alpha_{lj}^3{\gamma^{5/2}}},   \label{StressFreeGamma_l1} \\
B_2^j   + B_3^j  &= Ra \frac{ \mathcal{C}'_{11}(\alpha_{lj})}{3\,\alpha_{lj}^3},   \label{StressFree1_l1}
\end{align}
\end{linenomath}
while the normal stress continuity conditions \eqref{NormalStressGamma} and \eqref{NormalStress1} yield
\begin{linenomath}
\begin{align}
B_1^{j}  {\cal{P}^{-}}  + 6 \gamma \left[ B_2^{j}   +  \gamma^{-5}  B_3^{j}  \right] + \frac{3}{\gamma^{2}}    B_4^{j} 
&= \left[2 + \frac{\alpha_{lj}^2 \gamma^{2} }{2}  \right]  Ra\, \frac{\mathcal{C}'_{11}(\alpha_{1j} \gamma) }{\alpha_{1j}^3 \gamma^{\frac{3}{2}}}  \label{NormalStressGamma_l1} \\
- B_1^{j}  {\cal{P}^{+}}  + 6  \left[ B_2^{j}   +   B_3^{j}  \right] + {3}   B_4^{j} 
&= \left[2 + \frac{\alpha_{lj}^2 }{2}  \right]  Ra\, \frac{\mathcal{C}'_{11}(\alpha_{1j} ) }{\alpha_{1j}^3 }  \label{NormalStress1_l1}
\end{align}
\end{linenomath}
when $\mathcal{P}^-\ll 1$ and $\mathcal{P}^+\ll 1$.
Using Eqs. \eqref{StressFreeGamma_l1} and \eqref{StressFree1_l1}, the coefficients $B_2^{j}$ and $B_3^{j}$ can be eliminated from Eqs. \eqref{NormalStressGamma_l1} and \eqref{NormalStress1_l1}, which give
\begin{linenomath}
\begin{align}
\gamma^{2} B_1^{j}  {\cal{P}^{-}} + {3}    B_4^{j} 
&=   \frac{\gamma^{5/2}\, \mathcal{C}'_{11}(\alpha_{1j} \gamma) }{2\,\alpha_{1j} } Ra, \label{NormalStressGamma_l2} \\
- B_1^{j}  {\cal{P}^{+}}  + {3}   B_4^{j} 
&= \frac{\mathcal{C}'_{11}(\alpha_{1j} ) }{2\,\alpha_{1j} } Ra. \label{NormalStress1_l2}
\end{align}
\end{linenomath}
Noting that
\begin{linenomath}
\begin{align}
\mathcal{C}'_{11}(\alpha_{1j} )&=-\frac{2}{\pi \alpha_{1j}}\frac{J_{\frac{3}{2}}(\alpha_{1j}\gamma)}{J_{\frac{3}{2}}(\alpha_{1j})} \\
\mathcal{C}'_{11}(\alpha_{1j} \gamma)&=-\frac{2}{\pi \alpha_{1j}\gamma}
\end{align}
\end{linenomath}
\citep{Chandrasekhar1961}, solving Eqs. \eqref{NormalStressGamma_l2} and \eqref{NormalStress1_l2} yields
\begin{linenomath}
\begin{align}
B_1^{j} &= \frac{ {J_{\frac{3}{2}}(\alpha_{1j}\gamma)}/{J_{\frac{3}{2}}(\alpha_{1j})} - \gamma^{\frac{3}{2}} }{\pi \alpha_{1j}^2} \frac{Ra}{{\cal{P}^{+}} + \gamma^{2}  {\cal{P}^{-}} },  \\
B_4^{j} &= \left[ \frac{1- \gamma^{3/2}{J_{\frac{3}{2}}(\alpha_{1j})}/{J_{\frac{3}{2}}(\alpha_{1j}\gamma)}}{1 + \gamma^{2}  {\cal{P}^{-}}/{\cal{P}^{+}} } - 1 \right] \frac{1}{3\,\pi \alpha_{1j}^2 } \frac{J_{\frac{3}{2}}(\alpha_{1j}\gamma)}{J_{\frac{3}{2}}(\alpha_{1j})}Ra.
\end{align}
\end{linenomath}

It can be seen that $B_2^{j}$, $B_3^{j}$ and $B_4^{j}$ are all $\sim Ra$, while $B_1^{j}\sim Ra/({\cal{P}^{+}} + \gamma^{2}  {\cal{P}^{-}})$. 
In the limit of small $Ra$, $\mathcal{P}^-$, $\mathcal{P}^+$, but finite $Ra/({\cal{P}^{+}} + \gamma^{2}  {\cal{P}^{-}})$, we therefore have $B_1^{j}\gg(B_2^{j},B_3^{j},B_4^{j})$.
To a good approximation, $p_{1j}$ is then given (from Eq. \eqref{GeneralSolution_p}) by 
\begin{linenomath}
\begin{equation}
p_{1j} \simeq  B_1^{j} r  ,  \label{Solution_p_translatio} 
\end{equation}
\end{linenomath}
and the poloidal scalar of the first unstable mode is   
\begin{linenomath}
\begin{equation}
P = \sum_j A_{1j}p_{1j}(r) Y_1^0(\theta,\phi)  \simeq \left( \sum_j A_{1j} B_1^{j} \right) r\  Y_1^0(\theta,\phi),
\end{equation}
\end{linenomath}
which corresponds to a translational motion (it can be verified that a $l=1$ flow with $P \propto r$ corresponds to a flow with uniform velocity).

In the limit of small $Ra$, the characteristic equation \eqref{CharacteristicEquation} for $l=1$ now writes
\begin{linenomath}
\begin{equation}
\left|\left| N_{\frac{3}{2},k} \frac{\alpha_k^2}{4} \delta_{kj} - Q_{kj}  \right|\right| = 0,  \label{CharacteristicEquation2}
\end{equation}
\end{linenomath}
where
\begin{linenomath}
\begin{equation}
Q_{kj} = B_1^{j} \int_\gamma^1\!\! \mathcal{C}_{11}(\alpha_{1k} r)   r^{5/2} dr .  \label{Qkj_2}
\end{equation}
\end{linenomath}
Making use of recurrence relations of the Bessel functions \citep{abramovich1965}, we find that the integral in Eq. \eqref{Qkj_2} can be written as
\begin{linenomath}
\begin{equation}
\int_\gamma^1\!\! \mathcal{C}_{11}(\alpha_{1k} r)   r^{5/2} dr = \frac{2}{\pi \alpha_{1k}^2}  \left[   \dfrac{J_{\frac{3}{2}}(\alpha_{1k} \gamma)}{J_{\frac{3}{2}}(\alpha_{1k} )} - \gamma^{\frac{3}{2}} \right] ,
\end{equation}
\end{linenomath}
which allows to write $Q_{kj}$ as
\begin{linenomath}
\begin{equation}
Q_{kj} = \frac{2}{\pi^2 \alpha_{1j}^2 \alpha_{1k}^2}  \left[\dfrac{J_{\frac{3}{2}}(\alpha_{1j}\gamma)}{J_{\frac{3}{2}}(\alpha_{1j})} - \gamma^{\frac{3}{2}}\right] \left[   \dfrac{J_{\frac{3}{2}}(\alpha_{1k} \gamma)}{J_{\frac{3}{2}}(\alpha_{1k} )} - \gamma^{\frac{3}{2}} \right]  \frac{Ra}{{\cal{P}^{+}} + \gamma^{2}  {\cal{P}^{-}} } .  \label{Q_kj_3}
\end{equation}
\end{linenomath}

Now, rewriting Eq. \eqref{CharacteristicEquation2} as
\begin{linenomath}
\begin{equation}
\left|\left|    \delta_{kj} -   \frac{4}{\alpha_k^2\, N_{\frac{3}{2},k}}  Q_{kj} \right|\right| = 0 
\end{equation}
\end{linenomath}
and using Sylvester's determinant theorem, we find that
\begin{linenomath}
\begin{equation}
\left|\left|    \delta_{kj} -   \frac{4}{\alpha_k^2\, N_{\frac{3}{2},k}}  Q_{kj} \right|\right| = 1 - 4 \sum_{i=1}^\infty \frac{Q_{ii}}{\alpha_i^2 N_{\frac{3}{2},i}} = 0,  \label{CharacteristicEquation3}
\end{equation}
\end{linenomath}
from which, using Eq. \eqref{Q_kj_3}, we obtain the critical value of $Ra/(\mathcal{P}^++\gamma^2{\mathcal{P}^-})$:
\begin{linenomath}
\begin{equation}
\left(\frac{Ra}{{\cal{P}^{+}} + \gamma^{2}  {\cal{P}^{-}} } \right)_c   = \frac{1}{4} \left\{ \sum_{i=1}^\infty \frac{1}{\alpha_{1i}^4} \frac{ \left[\dfrac{J_{\frac{3}{2}}(\alpha_{1i}\gamma)}{J_{\frac{3}{2}}(\alpha_{1i})} - \gamma^{\frac{3}{2}}\right]^2}{    \left[ \dfrac{J_{\frac{3}{2}}^2(\alpha_{1i} \gamma)}{J_{\frac{3}{2}}^2(\alpha_{1i})} - 1  \right]}  \right\}^{-1} . \label{CriticalRaP}
\end{equation}
\end{linenomath}






\small








\end{document}